\documentclass{article}
\usepackage{epsfig}
\usepackage{epsf}

\def\slash#1{\setbox0=\hbox{$#1$}
   \dimen0=\wd0 \setbox1=\hbox{/} \dimen1=\wd1
   \ifdim\dimen0>\dimen1 \rlap{\hbox to \dimen0{\hfil/\hfil}} #1
   \else  \rlap{\hbox to \dimen1{\hfil$#1$\hfil}} / \fi}
   
\begin{document}

\title{Continuous local model \\ for two-dimensional spin ice}

\author{J. E. da Fonseca, C. W. Morais, \\ A. L. Mota{\footnote{motaal@ufsj.edu.br}}  and L. H. C. M. Nunes \\
\\ Departamento de Ci\^{e}ncias Naturais, \\ Universidade Federal de S\~{a}o Jo\~{a}o del Rei, \\
C.P. 110,  CEP 36301-160, S\~ao Jo\~ao del Rei, Brazil}

\maketitle

\begin{abstract}
We propose a classical continuous model Hamiltonian with a ground state presenting a spin ice structure. We analyze the introduction of metastable excitations on this ground state, showing the emergence of pairs of magnetic monopoles. The interaction between monopoles and dipoles in the system is studied. As a consequence, we obtain an effective nonlocal interaction between monopoles and dipoles from a local classical spin model.
\end{abstract}


%

\section{Introduction}

The emergence of magnetic monopoles in geometrically frustrated magnetic materials (spin ice)\cite{Castelnovo2008}  is one of the most surprising and interesting phenomena recently discovered, raising great interest to both experimental\cite{Bramwell2009,Fennel2009,Morris2009,Ladak2010,Mengotti2011,Giblin2011,Morgan2011,Farhan2016} and theoretical\cite{Jaubert2009,Mol2009,Castelnovo2010,Mol2010} research. Spin ice are magnetic materials with ferromagnetic interactions whose anisotropy jeopardize the alignment of all the spins, resulting in (degenerate) ground states with zero magnetization \cite{Harris1997}. These ground states obey the so called ice rule, the spins arranged in such a way that, on each vertex of the lattice, two of them point outward and the other two point inward the vertex, resulting in zero magnetic flux. It was predicted that excitations on the ground state of the spin ice, corresponding to the violation of the ice rule by spin inversion, lead to the emergence of quasi-particles with magnetic properties that justifies their denomination as magnetic monopoles \cite{Castelnovo2008, Silva2013}. 
\begin{figure}[!ht]
\begin{center}
\includegraphics[angle=0,width=0.60\columnwidth]{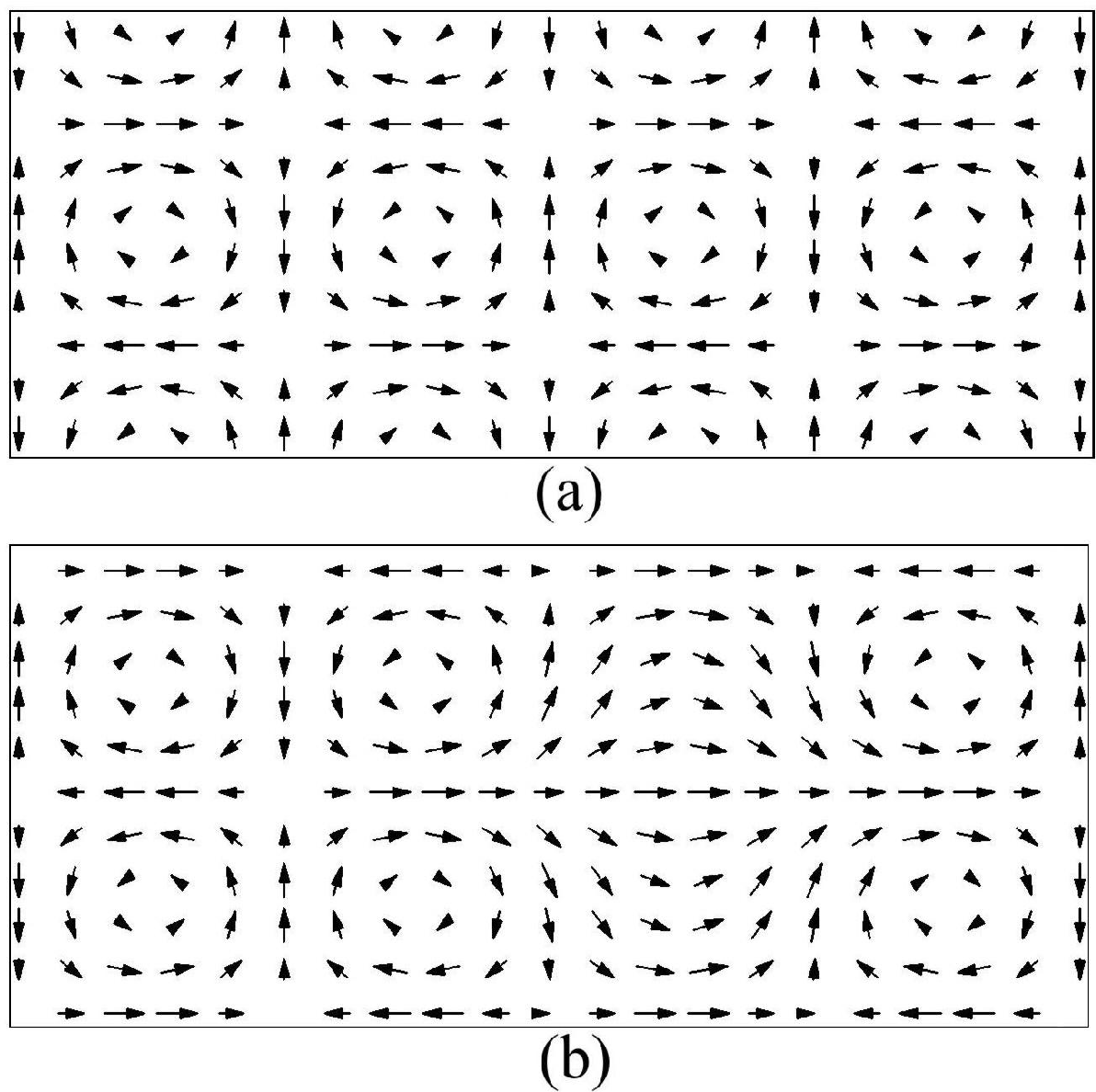}
\end{center}
\caption{Vector plot of the ground state (a) and excited state (b) of the model Hamiltonian.}
\label{groundstate}
\end{figure}

There are some characteristics of the spin ice systems that are essential for the emergence of magnetic monopoles in both three-dimensional (3D) real spin ice compounds, such as $Dy_2Ti_2O_7$ or $Ho_2Ti_2O_7$, or in artificial two-dimensional (2D) spin ice, as 2D square lattices of permalloy nano-islands \cite{Nisoli2013,Wang2006,Morgan2011} - the ice rule (divergenceless of the ground state), 
 periodicity (of the pyrochlore and artificial square lattices) and anisotropy (the magnetic moments are constrained to point along fix axes). Also in magnetic charged spin ice \cite{Ladak2010} (honeycomb spin ice), the last two characteristics are present.  In both 3D and 2D spin ice, the magnetic monopoles are joined by a string, a sequence of inverted spins. In 3D systems, the amount of energy necessary to separate apart two monopoles  is finite, leading to free monopoles, whereas in the 2D systems the monopoles are confined, the string energy grows with the string length \cite{Mol2009}. 
 
The thermodynamics properties of spin ice structures \cite{Farhan2013,Budrikis2012,Silva2012,Leon2016}, as well as the interaction between monopoles, has been successfully described by discrete Hamiltonians that take into account the local nearest neighbor interactions and non local (long range) dipolar interaction between the spins of the system \cite{Siddharthan1999,Ruff2005,Castelnovo2008,Mol2009}. Also, the magnetization dynamics in artificial spin ice under the action of an external magnetic field was recently studied in a discrete mesoscopic model\cite{Mellado2010}.
Here, we propose a new classical local model Hamiltonian for a continuous spin system inspired in the artificial two-dimensional spin ice. The ground state of the model can be mapped in the discrete artificial spin ice ground state. We will show that, as in the discrete case, the introduction of metastable excited states on the system causes the emergence of quasi-particles - the magnetic monopoles. We will analyze the stability of the excitations, the emergence of magnetic monopoles and the interactions between magnetic monopoles or dipoles in the system.

The model presented here is to be understood as an effective model for the 2D artificial spin ice systems. The correct choice of the model parameters can result, for example, in the same dependence of the system energy with magnetic monopoles separation as in the discrete model \cite{Mol2009}. There are several features of the continuous model (and its possible extension to 3D cases) that are useful to the study and understanding of spin ice - for example, analytical expressions for both ground and excited states can be very useful to the study of the dynamics and thermodynamics of the system. Also, there are several tools already developed to study the role of the symmetries (and their breakdown) of continuous Hamiltonians, and this can contribute to a better understanding of the phenomena involved. Nonetheless, it also suggests the possibility of experimental realizations of quasi-continuous spin ice systems - for example, by deposition of magnetic thin film over a patterned underlayer (an array of depressions with lattice parameter of the order of tenths of microns, but thickness of a few tens of nanometers), as already realized in some technological applications \cite{Arnold2004}. The shape anisotropy of the film deposited on the depressions can force the atomic spins to align lengthwise with the long axis, implementing the geometric anisotropy necessary to the appearance of frustration.

This paper is organized as follows: in Section 2 we present and discuss the model Hamiltonian and their ground and excited states. Numerical results for the parameters fitting and the condition for the stability of the excited states are presented in Section 3. The presence of the monopole-like magnetic excitations, strings and magnetic dipoles are discussed in Section 4. Finally, Section 5 brings our conclusions.

\section{The Model}
The model for a two-dimensional square latticed spin ice can be constructed from the ground state given by
\begin{equation}
\vec{S_0}(\vec{r})=-sin\Big(\frac{\pi x}{a}\Big) cos\Big(\frac{\pi y}{a}\Big)\vec i+sin\Big(\frac{\pi y}{a}\Big) cos\Big(\frac{\pi x}{a}\Big)\vec{j}, \label{S0}
\end{equation}
where $a$ is the lattice spacing. Throughout this paper, we will set $a=\pi$, without loss of generality. Figure 1(a) shows the mapping of the spin field, Eq.(\ref{S0}). 
At each vertex of the lattice ($x=n\pi, y=m\pi$, $n$ and $m$ integer) 
the ice rule is obeyed: each vertex has no magnetic charge (zero magnetic flux). The ground state satisfies the relations $\vec \nabla .\vec{S_0}=0$ and $\nabla^2 \vec S_0 = -2\vec S_0$.  

By introducing an auxiliary field $B=\frac{1}{2} (\frac{\partial S_{y}}{\partial x}-\frac{\partial S_{x}}{\partial y})$, we define our model by the Hamiltonian
\begin{eqnarray}
H &=& \int \int \Big\{ \kappa (\vec \nabla . \vec S)^2 + \lambda \Big((S_x- \frac{\partial B}{\partial y})^2+(S_y+\frac{\partial B}{\partial x})^2\Big)  \nonumber \\
&& + \mu \Big( |\vec S|^2 + B^2 - \frac{\partial S_x}{\partial x}\frac{\partial S_y}{\partial y} - 1 \Big)^2 \Big\} dx dy \label{hamiltonian}.
\end{eqnarray}
where $\kappa$, $\lambda$ and $\mu$ are free parameters of the model. From now on, we will assume $\kappa=1$, implying that $H$, $\lambda$ and $\mu$ are given in units of $\kappa$.

Since all terms of the Hamiltonian are positive-definite, the minimum of $H$ occurs only if each quadratic term on its integrand vanishes. So, the two first terms on the integrand on Eq.(\ref{hamiltonian}) implies that the ground state presents divergenceless and periodicity, as desired. Due to the presence of the product of derivatives $\frac{\partial S_x}{\partial x}\frac{\partial S_y}{\partial y}$ on the third term, the Hamiltonian is not invariant under continuous rotations in the $xy$ plane, although it is symmetric under discrete $\pi/2$ rotations, a feature that is shared by the actual artificial 2D spin ice system. So, this is the continuous version of the spin ice anisotropy. This third term is crucial to the model, since it fixes the amplitude of the ground state, so that solutions with low amplitudes do not imply, necessarily, in lower energies for the system. The ground state of Eq.(\ref{hamiltonian}) is infinitely degenerate, since any translation or discrete $\pi/2$ rotations on $\vec S_0$ imply also in a solution with zero energy. This is the manifestation of the spin ice residual entropy of the ground state in the continuous model.

Next, we introduce an excitation by adding to the ground state a localized function that ``flips'' the spins located around certain position $\vec r_e$. We propose $\vec\xi_{r_e}(\vec r)=e^{-\gamma(\vec r-\vec r_e)^2} f((\vec r-\vec r_e)^2)$, where the Gaussian function indicates that the excitation is localized, with $\gamma$ being the excitation width, and $f((\vec r-\vec r_e)^2) = A (\alpha (\vec r-\vec r_e)^4-2) (cos(\theta)\vec i+sin(\theta) \vec j)$, a function centered in $\vec r_e$. The free parameters $A$ (the excitation amplitude), $\alpha$, $\theta$ and $\gamma$ are to be fixed in order to minimize $H$. We will refer to $A$ as the amplitude of the excitation, $A=0$ implying that the system is in the ground state. 
The parameter $\theta$ gives the direction of the excitation, and was introduced in order to show that the excitation acts in the opposite direction of the ground state, locally flipping the spin. The functional form of the excitation $\vec \xi_{r_e}$ we employ here was chosen in order to contain the fewest possible free parameters, and it was tested against other functional forms with the same number of parameters. The fourth order power on $f$ was also tested against other even powers of $|\vec r-\vec r_e|$, and for the values of $\lambda$ and $\mu$ used here, showed up to be the best choice. 
Besides, we also found out that there are no qualitative (and very small quantitative) changes on the results by employing other even powers on $f$.
\begin{figure}[!ht]
\begin{center}
\includegraphics[angle=0,width=0.60\columnwidth]{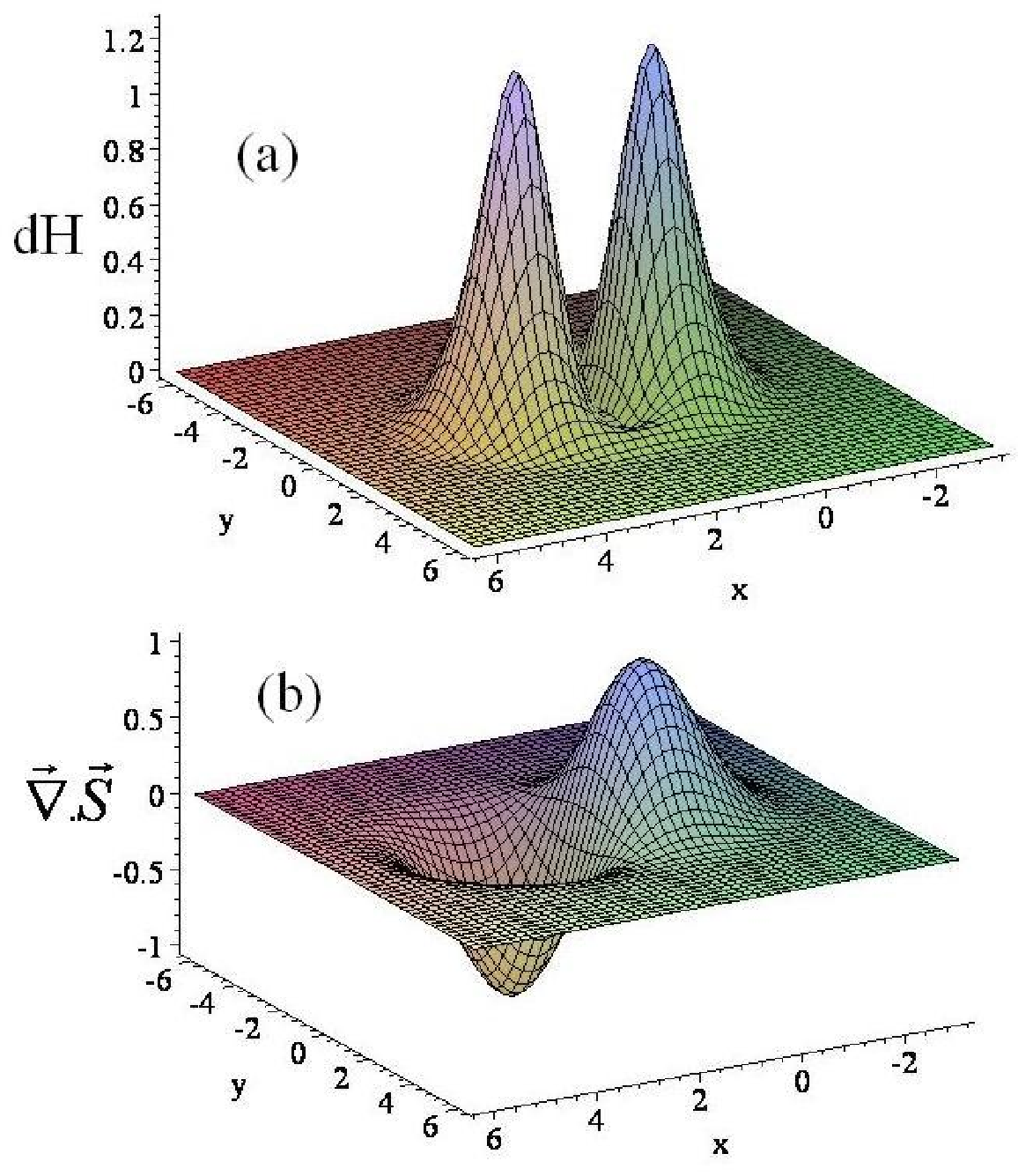}
\end{center}
\caption{The density of energy (a) and density of magnetic flux (b) of the system with the introduction of 1 excitation.}
\label{dH}
\end{figure}

\section{Numerical Results}
The relatively simple form of $\vec S_0$ and $\vec\xi_{r_e}$ allows, in principle, the Hamiltonian to be evaluated analytically. However, the number of terms involved are of the order of $500$, and even prohibitively higher when a larger number of excitations is introduced. Fortunately, the localized nature and smoothness of the excitation allows one to construct accurate algorithms for the evaluation of $H$. Here, we used the composite Simpson's rule for two-dimensional integrals with $500 \times 500$ initial sampled points, and convergence in the 8th decimal place. The estimated maximum error is in the 5th decimal place (typically the 7th significant figure). For optimizing the free parameters of the excitation, we employed the steepest descent algorithm, with convergence for the energy in the 8th decimal place.

Figure 1(b) shows the mapping of an excited state 
for $\alpha = 5.419 \times 10^{-2}$, $\gamma = 0.3426$, $\theta=\pi$ and $A=1$. The parameters were obtained for $\lambda = 10^{-6}$ and $\mu=2.6$ (the small value of $\lambda$ was chosen in order to minimize the string tension). The excitation flips the spins only in a localized region, as expected, and the two adjacent vertexes no longer obey the ice rule. 
As in the discrete lattice, this corresponds to the emergence of the magnetic monopoles - Figure 2(a) shows the density of Hamiltonian at the vicinity of the excitation. One can clearly distinguish two localized distributions of energy, corresponding to two quasi-particles in the system
with opposite magnetic charge (see the density of flux diagram, Fig. 2(b)).
\begin{figure}[!ht]
\begin{center}
\includegraphics[angle=0,width=0.60\columnwidth]{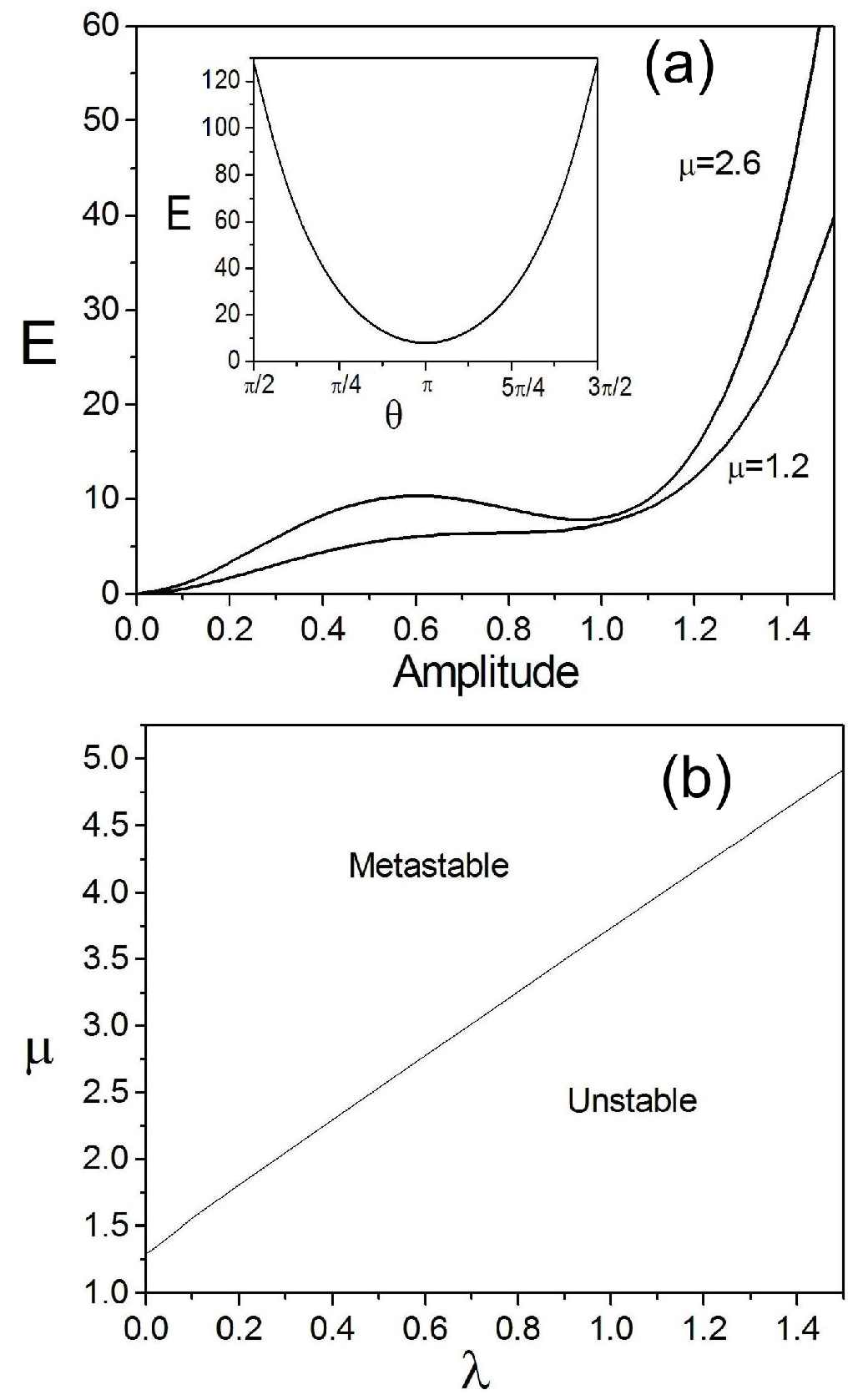}
\end{center}
\caption{ (a) The system energy as a function of the excitation amplitude $A$. 
For $\mu > \mu_c$, there is a local minimum at $A \approx 1$. On the inset - The energy as a function of the excitation direction. (b) The metastability diagram of the excited state.} 
\label{Stability}
\end{figure}

The results are, obviously, dependent on the Hamiltonian coefficients $\lambda$ and $\mu$. For $\mu=0$, a quick evaluation of Eq.(\ref{hamiltonian}) shows that the excitation is not stable - the energy monotonically goes down to zero as $A \rightarrow 0$. For values of $\mu$ higher than certain value $\mu_c$, the energy has a local minimum for some $A \ne 0$. This can be seen in Figure 3(a) - for $\lambda = 10^{-6}$ and $\mu = 1.2$, the energy of the system presents only one stable minimum at the ground state. For $\mu=2.6$, however, there is a local minimum at $A \approx 1$. So, above $\mu_c$ the system presents metastable excited states. 
The inset in Figure 3(a) shows the stability of the $\mu=2.6$ excited state with respect to the direction of the excitation - the minimum energy occurs for $\theta = \pi$, opposite to the direction of $\vec S_0$ at $\vec r_e$. Figure 3(b) shows that the $\mu \times \lambda$ plane is split into two regions, above $\mu_c$ the excitation corresponds to a metastable state.

The model Hamiltonian for a continuous spin system, Eq.(\ref{hamiltonian}), presenting a non-trivial ground state that satisfies the spin ice rule, together with the stability of the excited state, are the key results of this paper. As a consequence, there is the emergence of the magnetic monopoles as localized energy distributions with opposite magnetic flux. 
It is worth to emphasize that this is the first time, to our knowledge, that a continuous Hamiltonian with a generalized spin ice ground state, without applying any boundary condition or external field, is presented. The presence of metastable excitations leading to localized energy distributions that interact with each other as quasi-particles, generating nonlocal effects from a purely local Hamiltonian model is also a highly nontrivial result.
In what follows, we shall see that, as in the discrete 2D artificial spin ice case, the energy of the system grows with the monopoles separation. We will also show that , due to the overlapping of the excitations, the model presents dipoles interaction. 

\begin{figure}[!ht]
\begin{center}
\includegraphics[angle=0,width=0.60\columnwidth]{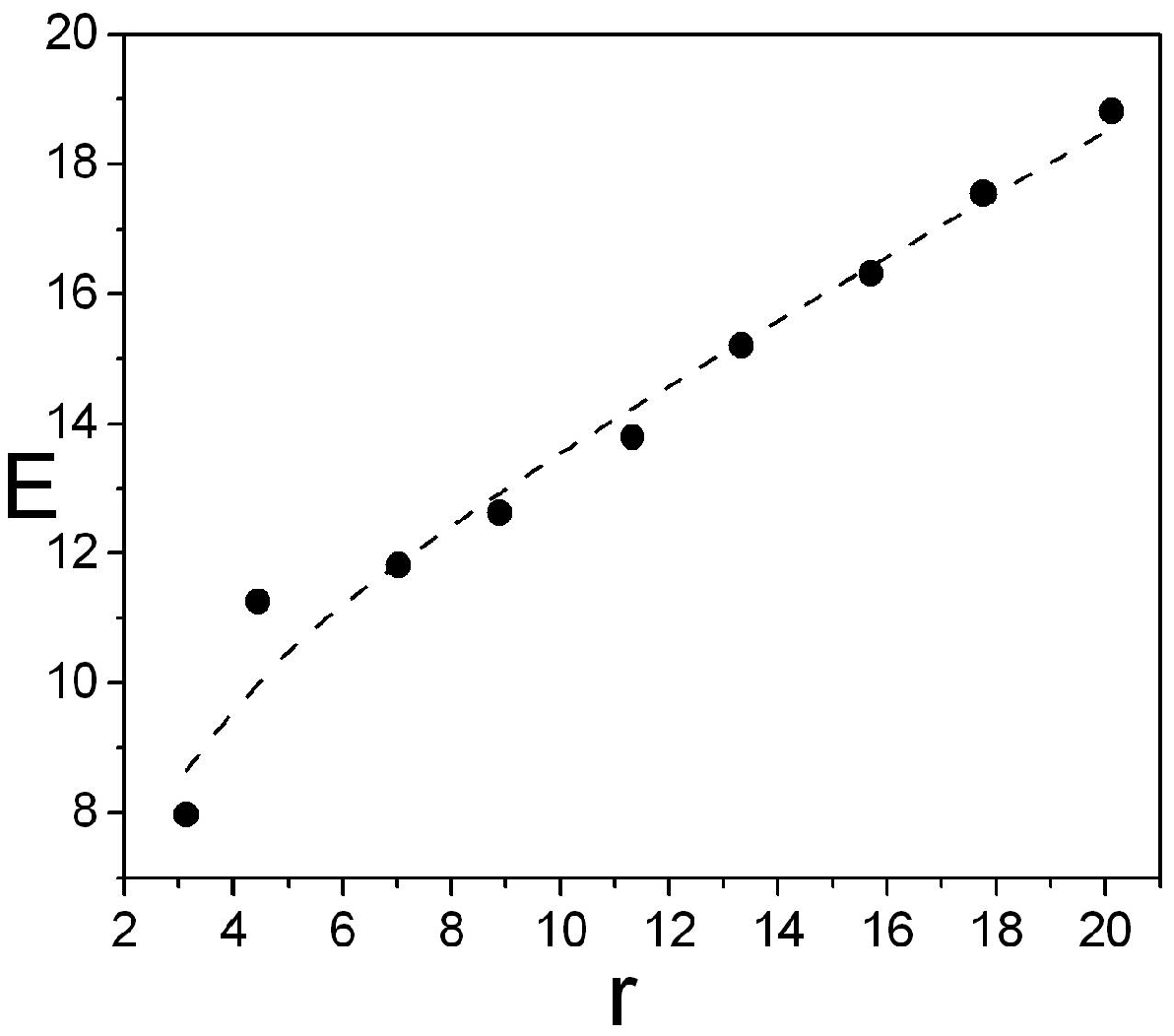}
\end{center}
\caption{The system energy as a function of the monopoles separation.}
\label{phasediag}
\end{figure}
\begin{figure}[!ht]
\begin{center}
\includegraphics[angle=0,width=0.70\columnwidth]{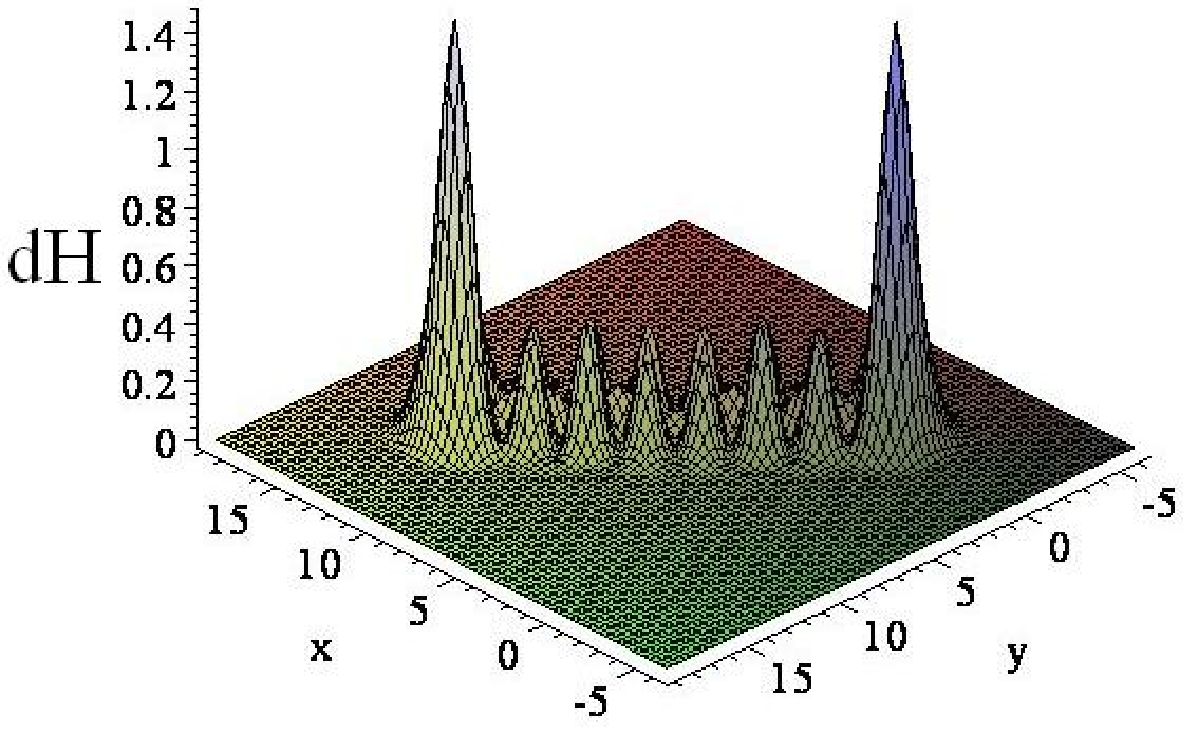}
\end{center}
\caption{The density of energy for 8 excitations introduced in line. The two higher peaks correspond to the two monopoles. The small oscillations corresponds to the string joining the two monopoles.}
\label{dH8exc}
\end{figure}

\section{Monopoles and Dipoles}

The introduction of a sequence of $N$ excitations in the system at subsequent positions $\{\vec r_i\}$, spatially separating the monopoles, is obtained by adding to the ground state the function $\Sigma_{i=1}^N \vec \xi_i(\vec r)$, where $\xi_i(\vec r) = e^{-\gamma(\vec r-\vec r_i)^2} (\alpha_i (\vec r-\vec r_i)^4-2) (cos(\theta_i)\vec i+sin(\theta_i) \vec j)$. We are already assuming $A=1$ and $\theta_i$ giving the opposite direction of the field $\vec S_0(\vec r_i)$. The free parameters to be optimized are $\gamma_i$ and $\alpha_i$, $i=1..N$. Symmetry can be used to reduce the number of parameters, so, for example, for the introduction of two adjacent excitations, only one set of parameters is necessary. 
For a large enough in-line chain of excitations, the parameters for the excitations at the middle of the chain do not differs significantly from each other, so, in general, only four sets of parameters are necessary. 
As in real spin ice systems, this string is not unique, and, as in Refs.\cite{Castelnovo2008,Mol2009}, the results presented here are for the shortest string connecting the two monopoles.

Fig. 4 shows the energy of the system as a function of the monopoles separation $r$. 
The energy grows with $r$, an evidence of the presence of the string joining the monopoles, as in the discrete case \cite{Mol2009}. From $N \geq 4$ approximately, the energy grows quite linearly with $r$. As a consequence, the monopoles are confined, i.e., from a specific $r$ value it would be energetically more favorable to disrupt the string in order to form two new isolated dipoles. However, at zero temperature the strings are stable - there are no fluctuations that could bring the large dipole configuration to the two smaller dipoles one. In actual 2D spin ice systems, the temperature necessary to flip a magnetic domain is prohibitively high, so there is no possibility of such a thermodynamical rupture to occur \cite{Wang2006}. However, a stochastic external magnetic field \cite{Cowburn2002, Ke2008, Mellado2010} could provide the fluctuations necessary to the create an effective thermal ensemble, allowing the string rupture. 

At small $r$, the behavior of the curve indicates that negative powers of $r$ should dominate in this region, as also observed in ref. \cite{Mol2009}. The dashed line in Fig. 4 represents the least square fitting of the energy to the function $\alpha + \beta r + \delta/r$. The fitting clearly shows that there are contributions of negative powers of $r$. The density of Hamiltonian  for a specific configuration with 8 excitations is presented in Figure 5. The two higher peaks correspond to the magnetic monopoles, joined by a ``string'' of small fluctuations of the energy.

By introducing two different chains of excitations of fixed length (two emergent dipoles) we can also evaluate the energy behavior as we separate the dipoles. The energy rapidly drops down to the noninteracting dipoles value ($2 E_0= 2 \times 11.25188$), since the excitations are characterized by a short excitation width $\gamma \approx 0.37$. We evaluate the energy of a configuration of two dipoles of length $\sqrt{2} \pi$ separated by the distances $r=\pi$ and $r=2\pi$ in the $y$ direction. The excitations are introduced at $\vec r_{11}=(\pi/2)\vec i$, $\vec r_{12}=\pi \vec i + (\pi/2)\vec j$,
$\vec r_{21}=(\pi/2) \vec i + r \vec j$ and $\vec r_{22}=\pi \vec i + (r + \pi/2) \vec j$. 
The $r=\pi$ configuration corresponds to two anti-parallel dipoles, the configuration energy results $E=19.54$, while for $r=2\pi$ the dipoles are parallel, and the configuration energy is $E=22.65$. As expected for a dipole interaction, it is attractive when the dipoles are anti-parallel ($E < 2E_0$) and repulsive when the dipoles are parallel ($E>2E_0$). However, it is important to note that the interaction between dipoles depends on the specific path employed to construct the strings, leading to a very rich phenomena to be studied. 
The short range dipoles interaction can be observed, as the configuration energy for $r=3\pi$ and $r=4\pi$ both agree to $2E_0$ at the 3rd decimal place.

The action of an uniform external magnetic field on the ground state and excited configurations can be easily evaluated by adding the interaction Hamiltonian $H_h = -\vec h. \int \vec S dx dy$ to  Eq.(\ref{hamiltonian}). With a not strong enough $\vec h$, in order to not change the system configuration, the evaluation of $H_h$ in the ground state reveals that there is no changes on the system energy - there is no monopoles to couple with the external field. With the external field aligned on the direction of $\vec S$ at the vicinity of an  excitation, the system energy is lowered, whereas with the external field on the opposite direction, the energy is increased, as can be easily inferred from the observation of the excited state configuration, Fig. 1(b). So there is an effective coupling of the magnetic monopoles with the external magnetic field. An interesting study to be addressed is how the external field can (1) provide transitions from the ground state to other phases, and (2) influence the stability of the excited solutions.

\section{Conclusion}
We have presented a classical continuous model whose ground state presents 2D spin ice structure. We have studied the metastability of the excitations introduced in order to locally flip the spins, with the consequent emergence of magnetic monopoles. The effective interaction between monopoles or dipoles was analyzed. As a general conclusion, we obtain effective non-locality (among monopoles and dipoles) from a classical local model. Further developments of the model presented here are its natural extension to 3D spin ice systems and the dynamics and thermodynamics of the model. 

\section{Acknowledgments}
We thanks A. R. Pereira and R. S. Freitas for useful discussions. We also acknowledge the financial support from CAPES, CNPq and FAPEMIG (Brazilian agencies).

\end{document}